\documentclass[prl,aps,floats,showpacs,superscriptaddress,twocolumn]{revtex4}

\usepackage{psfig}
\usepackage{bm}

\begin{document}

\title{Quantum cloning at the light-atoms interface: \\
copying a coherent light state into two atomic quantum memories}

\author{Jarom\'{\i}r Fiur\'{a}\v{s}ek}
\affiliation{QUIC, Ecole Polytechnique, CP 165,
Universit\'{e} Libre de Bruxelles, 1050 Bruxelles, Belgium }
\affiliation{Department of Optics, Palack\'{y} University,
17. listopadu 50, 77200 Olomouc, Czech Republic}

\author{Nicolas J. Cerf\,}
\affiliation{QUIC, Ecole Polytechnique, CP 165,
Universit\'{e} Libre de Bruxelles, 1050 Bruxelles, Belgium }

\author{Eugene S. Polzik}
\affiliation{QUANTOP, Niels Bohr Institute, Copenhagen University, 
Blegdamsvej 17, DK-2100 K{\o}benhavn, Denmark}

\begin{abstract}
A scheme for the optimal Gaussian cloning of coherent light states 
at the light-atoms interface is proposed. The distinct feature of this proposal
is that the clones are stored in an atomic quantum memory, which
is important for applications in quantum communication. The atomic
quantum cloning machine requires only a single passage of the light pulse 
through the atomic ensembles followed by the measurement of a light quadrature
and an appropriate feedback, which renders the protocol experimentally
feasible. An alternative protocol, where one of the clones is carried by the
outgoing light pulse, is discussed in connection with quantum key distribution.
\end{abstract}

\maketitle

Quantum information processing with continuous variables provides an interesting
alternative to the traditional qubit-based approach. Continuous
variables (CV) seem to be particularly suitable for quantum communication
applications, as for example quantum teleportation \cite{Furusawa98}
or quantum key distribution (QKD) \cite{Grosshans03}.
Another important feature of CV is the feasibility 
of the light-atoms quantum interface \cite{Kuzmich03,Julsgaard01}, 
which unlike its qubit analogue does 
not require strongly coupled cavity QED regime for deterministic operations. 
Along these lines, the prospect of developing a quantum memory
for light with macroscopic atomic ensembles has been explored 
\cite{Hald,Kuzmich00,Schori,Kozhekin}. Such a quantum memory
is crucial for applications such as quantum repeaters or quantum secret sharing.
\par

In this paper, we show that the optimal Gaussian cloning \cite{Cerf-book,Cerf00}
of a coherent state of a traveling light beam
can be achieved via its off-resonant interaction with atomic ensembles.
In the envisaged experiment,
the light beam interacts with two atomic ensembles A and B (see Fig.~1).
The two resulting (approximate) clones are stored in the quantum states
of the collective atomic spins of the clouds A and B;
we thus achieve cloning into an atomic quantum memory. This is fairly
distinct from the all-optical setup for CV quantum cloning based on the use 
of a parametric optical amplifier \cite{Braunstein01,Fiurasek01}.
A variation of our approach allows one of the clones to be
stored in the atomic cloud, the second one being carried by the outgoing light
pulse (see Fig.~3). This feature makes this second scheme 
particularly attractive for eavesdropping on the QKD schemes 
utilizing coherent states \cite{Grosshans03}.
An eavesdropper, Eve, who intercepts the quantum signal may keep 
one clone in the atomic memory and send the other one down the line. 
Eve waits until the receiver announces
the measurement basis ($x$ or $p$ quadrature),
and only then performs the corresponding measurement on her clone.
Note that such Gaussian cloning attacks are the optimal finite-size
attacks on a certain class of QKD schemes with coherent states \cite{Grosshans04}.
\par

\begin{figure}[!b!]
\centerline{\psfig{figure=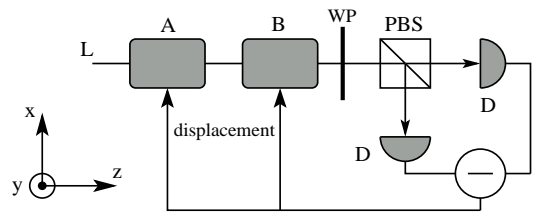,width=0.80\linewidth}}
\caption{Setup for the CV cloning of light 
into an atomic quantum memory. A light beam $L$ polarized along
the $x$-axis and propagating along the $z$-axis passes through two atomic
ensembles $A$ and $B$ polarized in the $x$ direction.
The $z$-component of the Stokes vector of the output light beam is measured
and the atomic states are then displaced accordingly.
The clones are stored in the atomic ensembles $A$~and~$B$.}
\end{figure}

The off-resonant interaction of light with an atomic ensemble can be described
by the effective unitary evolution operator $U=\exp(-iaS_z J_z)$,
where $S_z$ denotes the $z$ component of the Stokes operator $\bm{S}$
describing the
polarization state of light, and $J_z$ stands for the $z$ component of the
collective atomic spin operator ${\bm J}$ \cite{Kuzmich98,Kuzmich00}.
The elements of the vectors $\bm{S}$ and $\bm{J}$ satisfy the standard angular
momentum commutation relations,
$[S_i,S_j]=i\epsilon_{ijk} S_k$ and $[J_i,J_j]=i\epsilon_{ijk} J_k$.
The effective coupling strength $a$ depends on the details of the level
structure of the atoms, the detuning between light and the atomic transition, 
and the geometry of the experiment \cite{Duan00}.
\par

Consider the situation where the light beam contains a strong coherent component
linearly polarized in the $x$ direction and, similarly, the atomic spins are
polarized along the $x$ axis. In this case, the mean values of
$S_x$ and $J_x$ attain macroscopic values $\langle S_x\rangle\approx N_L/2$
and $\langle J_x\rangle\approx N_A/2$, where $N_L$ and $N_A$ denote the number
of photons in the light beam and the number of atoms in the ensemble,
respectively. This implies that we may approximate the operators $S_x$ and
$J_x$ by their mean values. It follows from the commutation relations
for $\bm{S}$ and $\bm{J}$ that the properly rescaled $y$ and $z$ components
of the vectors $\bm{S}$ and $\bm{J}$ satisfy
the canonical commutation relations for two conjugate quadrature operators
$x$ and $p$, namely $[x,p]=i$. We can thus introduce the effective
quadrature operators for light and atoms:
$x_L=S_y/\sqrt{\langle S_x\rangle}$, $p_L=S_z/\sqrt{\langle S_x\rangle},$
$x_A=J_y/\sqrt{\langle J_x\rangle}$, and $p_A=J_z/\sqrt{\langle J_x\rangle}.$
Note that $x_L$ and $p_L$ can be interpreted as the quadratures of the
optical mode linearly polarized in the $y$ direction.
This correspondence forms the basis for the implementation of
continuous-variable quantum information processing
using the off-resonant coupling between light and atoms.
If we rewrite the unitary transformation $U$ in terms of the quadrature
operators, we get
\begin{equation}
U=\exp(-i \kappa\, p_L\, p_A),
\end{equation}
where $\kappa=a\sqrt{N_L N_A}/2$.
Since the effective Hamiltonian generating $U$ is quadratic, i.e.,
$H=p_L\, p_A$, $U$ is a linear canonical transformation of the quadrature 
operators. By applying local phase shifts to the light and atoms
(i.e. by rotating the vectors $\mathbf{S}$ and $\mathbf{J}$) we may modify the
effective Hamiltonian to $H=x_L\, p_A$ or $H=p_L\, x_A$.
The rotation of the polarization of light is easily
performed by sending the light pulse through waveplates. The polarization
of the atomic cloud can be rotated by applying strongly detuned
classical laser pulses \cite{Kuzmich00,Duan00}.
\par

\paragraph{Two-pass atomic quantum cloning.}

Consider the effective unitary transformation $U=\exp(-i\kappa\, x_L\, p_A)$.
In the Heisenberg picture, the quadratures evolve according to
\begin{equation}
\begin{array}{lcl}
x_L^{out}=x_L^{in}, &\qquad& x_A^{out}=x_A^{in}+\kappa\, x_L^{in}, \\
p_{L}^{out}=p_L^{in}-\kappa\, p_{A}^{in}, & & p_{A}^{out}=p_A^{in}.
\end{array}
\label{inout}
\end{equation}
Specifically, when $\kappa=1$, we obtain the so-called continuous-variable
controlled-NOT (C-NOT) gate, i.e. 
$|x\rangle_L|y\rangle_A \rightarrow |x\rangle_L |y+x\rangle_{A}$,
where $|x\rangle$ and $|y\rangle$ represent the eigenstates of the
$x$ quadratures. The mode $L$ is the control mode
while $A$ is the target mode. Since
the CV C-NOT is not its own inverse, we also introduce C-NOT$^\dagger$,
which is obtained by choosing $\kappa=-1$ in Eq.~(\ref{inout}).
As pointed out in \cite{Cerf00}, it is possible to construct
the optimal Gaussian cloning machine with a quantum network made of four
such CV C-NOT gates if some particular
ancillary state can be prepared. 
\par

\begin{figure}
\centerline{\psfig{figure=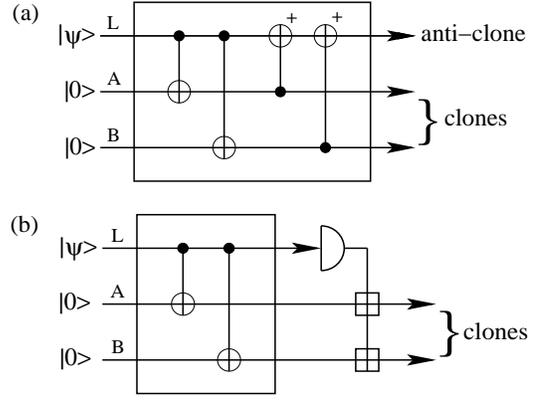,width=0.8\linewidth}}

\caption{(a) Network for optimal symmetric Gaussian cloning of coherent
states. The lines indicate the C-NOT gates, $\bullet$ labels control mode and
$\oplus$ indicates target mode. (b) Simplified cloning network where two
C-NOT gates are replaced by the measurement of the $p_L$ quadrature followed by
displacements of the atomic $p$ quadratures, as schematically indicated
by the open boxes.}
\end{figure}

The cloning network that we propose is depicted in Fig. 2(a).
The three relevant modes labeled $L$, $A$, and $B$ correspond,
respectively, to the light beam
and the two atomic ensembles (see Fig. 1). The atomic
ensembles are initially prepared in the vacuum state and
the light beam carries the coherent state to be cloned. The cloning can be
divided into two steps. First, the information about the $x$ quadrature
is transferred from the light into the atomic samples by applying
two C-NOT gates where the light is the control mode and the atomic modes
are the targets. After this first passage of light, the quadratures evolve as
\begin{equation}
\begin{array}{lll}
x_A^\prime=x_A^{in}+x_L^{in}, &\qquad & p_A^\prime=p_A^{in}, \\
x_B^\prime=x_B^{in}+x_L^{in}, & & p_B^\prime=p_B^{in}, \\
x_L^\prime=x_L^{in}, & & p_L^\prime=p_L^{in}-p_A^{in}-p_B^{in}.
\end{array}
\label{stepone}
\end{equation}
In the second step, the information about the $p_L$ quadrature is transmitted.
This is accomplished by two C-NOT$^\dagger$ gates
where now the atomic modes play the role of the controls and the light
is the target (the reverse information transfer of $p$ from light to the atomic
samples works by back-action). The output quadratures can thus be expressed
in terms of the input ones as
\begin{equation}
\begin{array}{lll}
x_A^{out}=x_A^{in}+x_L^{in}, &\qquad & p_A^{out}=p_L^{in}-p_B^{in}, \\
x_B^{out}=x_B^{in}+x_L^{in}, & & p_B^{out}=p_L^{in}-p_A^{in}, \\
x_L^{out}=-x_L^{in}-x_A^{in}-x_B^{in}, & &
p_L^{out}=p_L^{in}-p_A^{in}-p_B^{in}.
\end{array}
\label{steptwo}
\end{equation}
which is the desired optimal Gaussian cloning transformation.
To illustrate this, consider the state of a single clone, say $A$.
For input coherent state $|\alpha\rangle_L$ in the optical mode $L$,
the atomic mode $A$  ends up in a mixed Gaussian state, 
namely the input coherent state with superimposed thermal noise.
The coherent component of clone $A$ is equal to
$\alpha$, which guarantees that all coherent states are cloned with the same
fidelity $F$. The latter is related to the mean number $\bar{n}$ of
thermal photons in $A$ by the formula \cite{Fiurasek01}
\begin{equation}
F= \frac{1}{\bar{n}+1}.
\label{Fidelity}
\end{equation}
On inserting $\bar{n}=1/2$ into Eq.~(\ref{Fidelity}) we obtain
$F=2/3$, which is the maximal fidelity achievable 
by Gaussian cloning machines \cite{Iblisdir00}.
Hence, the proposed cloning procedure is optimal.
\par

\paragraph{Single-pass atomic quantum cloning.}
Let us now consider the practical
realization of this procedure in the specific system illustrated in Fig.~1.
The transformation (\ref{stepone})
is accomplished by the passage of the light pulse through both atomic samples
$A$ and $B$ with the polarization settings chosen such that the
effective coupling between the light and the atomic samples
is described by the Hamiltonian $H_1=x_L(p_A+p_B)$.
In practice, the pulse is very long compared to
the distance between the atomic samples, so the light interacts
simultaneously with both samples $A$ and $B$.
\par

The realization of the last two C-NOT$^\dagger$ gates, however, brings
complications. The light pulse should pass for a second time through
the atomic samples, and, before this second passage,
the polarization of the light and the atoms should be rotated to switch
the effective coupling to $H_2=-p_L(x_A+x_B)$. This may be very
difficult to accomplish because, in the current experiments, the pulse
is several hundred kilometers long.
The leading part of the pulse, which already traveled through
the atomic ensembles, would have to be stored until the tail of the
pulse also passes through the atoms. Only then can the pulse be fed back
into the input. Luckily, this complicated and technically challenging
procedure can be avoided because the last two C-NOT$^\dagger$ gates
can be replaced by a measurement of the $z$-component of the Stokes vector
of light followed by some appropriate displacement of the atomic quadratures. The latter task can be accomplished by a
tiny rotation of the polarization state of the collective atomic spin
\cite{Kuzmich00,Duan00}.
\par

This crucial simplification renders our proposal
experimentally feasible. The simplified cloning procedure is
illustrated in Fig. 2(b). The measurement of $S_z$ is equivalent to the
measurement of $p_{L}^\prime$. Once the classical measurement outcome
$p_L^{cl}$ is known, one has to displace the atomic $p$ quadratures
of ensembles $A$ and $B$ as follows
\begin{equation}
p_A^\prime\rightarrow p_A^\prime+p_L^{cl}, \qquad
p_B^\prime\rightarrow p_B^\prime+p_L^{cl}.
\end{equation}
It is immediate to see that, after displacement, the resulting quadratures
of the atomic modes are equal to those given in Eq.~(\ref{steptwo}), 
hence the optimal cloning is achieved with a single passage of light.
\par

\paragraph{Cloning into atoms and light.}
The above protocol produces two clones stored in two atomic memories
$A$ and $B$. If the cloning is used as an eavesdropping attack, then Eve would
like to store her clone in the memory while the second clone should be sent
as a light pulse down the communication line. 
One option would be to transfer one of the
clones from the atomic memory back to light. Such a procedure, however, would
require either strong entanglement \cite{Kuzmich00}
or several passages of the light pulse
through the atomic ensemble \cite{Kuzmich03,Fiurasek03},
which is currently infeasible.
Instead, one can use the scheme depicted in Fig. 2(b) and replace the
second atomic memory with a light beam $B$.
The required QND-type interaction between the two
light beams $L$ and $B$ could be realized with the use of a non-degenerate
optical parametric amplifier placed in between two unbalanced beam splitters
\cite{Pereira94,Bencheikh95}. After the measurement of the
quadrature $p_L^\prime$, the light quadrature $p_B$ should be displaced
similarly as in the CV teleportation experiments \cite{Furusawa98}.
\par

\begin{figure}
\centerline{\psfig{figure=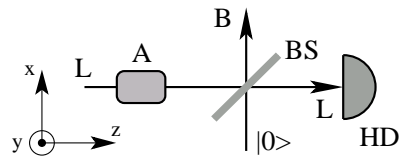,width=0.6\linewidth}}
\caption{Setup that produces one clone stored in an
atomic quantum memory $A$ while the second ``flying'' clone is carried by
the light beam $B$. BS is a balanced beam splitter and HD denotes homodyne
detector. The clone in the light beam $B$ is squeezed in the $x$ 
quadrature by a factor of $\sqrt{2}$.}
\end{figure}

An even simpler alternative scheme is shown in Fig.~3. 
Here, the light beam $L$
passes through a single atomic ensemble $A$ ($H=x_L\, p_A$, $\kappa=1$)
and then impinges on a balanced
beam splitter BS whose second input port is in the vacuum state.
Note that $\langle S_{x,L}^{out}\rangle=\langle S_{x,B}^{out}\rangle=\langle
S_{x,L}^{in}\rangle /2$ and all output quadratures are defined
as properly normalized $y$ and $z$ components of the Stokes vectors 
such that canonical commutation relations are satisfied.
The quadrature $p_L$ of the output beam $L$
(i.e., the $S_{z,L}$ component of the Stokes vector)
is measured by the homodyne detector HD, and the quadratures $p_A$ and
$p_B$ are displaced according to
\begin{equation}
p_A\rightarrow p_A+\sqrt{2}p_L, \qquad p_B\rightarrow p_{B}+p_{L}.
\end{equation}
It is easy to show that the resulting output quadratures of the modes 
$A$ and $B$ can be expressed in terms of the input quadratures as
\begin{equation}
\begin{array}{lcl}
x_{A}^{out}=x_L^{in}+x_A^{in}, & \quad &
x_B^{out}=\frac{1}{\sqrt{2}}(x_L^{in}+x_B^{in}), \\
p_{A}^{out}=p_L^{in}-p_B^{in}, & \quad & p_B^{out}=\sqrt{2}(p_L^{in}-p_A^{in}).
\end{array}
\end{equation}
The atomic memory $A$ contains one clone, while the
light beam $B$ contains the other clone squeezed in the $x$ quadrature
by a factor of $\sqrt{2}$. To restore the optimal clone in the light
mode $B$ one would have to ``unsqueeze'' this beam using a phase-sensitive
(degenerate) optical parametric amplifier with squeezing
factor $\sqrt{2}$.
\par

\paragraph{Asymmetric cloning.}
Let us now demonstrate how to make the cloning machine asymmetric.
This is particularly interesting in the context of
quantum cryptography where it enables Eve to choose a trade-off
between the quality of her copy (hence, the information she can extract
from it) and the unavoidable noise that is added to the copy sent
to the receiver. 
We  pursue an approach inspired from \cite{Cerf00}, which relies on 
the ``preprocessing'' of the initial states of the atomic modes. Suppose 
that the atomic modes $A$ and $B$ are both initially prepared
in a pure single-mode 
squeezed vacuum state. Mode $A$ is squeezed in the $x$ quadrature
and mode $B$ is squeezed in the $p$ quadrature, and the squeezed variance
is $V$ in both cases.
The scheme depicted in Fig. 2(b) then produces two asymmetric clones whose
fidelities read $F_A=1/(1+V)$ and $F_B=4V/(4V+1)$, in accordance with
\cite{Fiurasek01}. This confirms that the device indeed 
realizes the optimal asymmetric Gaussian cloning of coherent states.
\par

The atomic ensembles can be squeezed using the same procedure that
was employed to entangle two atomic ensembles \cite{Duan00,Julsgaard01}.
Coherent light pulses are used to 
perform  quantum nondemolition measurements of two commuting nonlocal
atomic quadratures $x_A+p_B$ and $x_A-p_B$. If the interaction
strength is $\kappa=\sqrt{1/(4V)-1/2}$, this projects 
the atoms onto squeezed states with squeezed variance $V$ and
a coherent component determined by the measurement outcome.
By applying a properly chosen displacement, we can set the coherent component
to zero, which results in a deterministic preparation of a pure
single-mode squeezed vacuum state of each ensemble A and B. 
Note that this procedure requires only a single 
passage of each light pulse through the atomic ensembles.
\par

\paragraph{Experimental feasibility.}
The feasibility of the experimental realization is basically determined by the
relative easiness of fulfilling the condition on the coupling constant 
$\kappa=a\sqrt{N_L N_A}/2=1$.
For room temperature atomic vapors, this value has been already achieved 
and even surpassed in \cite{Julsgaard01}. For cold atomic vapor, the value 
of $\kappa=1$ is also well within reach. The coupling constant can be 
rewritten as $\kappa=(\sigma\gamma/A\delta)\sqrt{N_L N_A}/2$ with 
$\sigma=(\lambda)^2/2\pi$, $\gamma$ - natural width of the transition 
and $\delta$ - detuning of the light from the transition. It is important 
to reduce the probability $\eta$ of the spontaneous emission, caused by 
the probe, so that this source of decoherence does not affect the cloning 
protocol. It has been shown \cite{Hamerrer} that spontaneous emission 
can be neglected if $\eta \ll 1/(1+\kappa^2)$. This condition translates 
into a usual for the free space coupling protocols condition on the optical 
density of the atomic sample,  $\alpha \gg 1$, since $\kappa^2=\alpha\eta$. 
This condition can be easily met with dense cold atomic samples.
\par

Ground magnetic or hyperfine states of alkali atoms are good candidates 
for a two level atomic system for cloning. Entanglement life time for such 
a system of up to 1 msec has been demonstrated in \cite{Julsgaard01}. 
Cold atoms can have the coherence life time of up to seconds, so that the 
life time of atomic clones can, in principle, be of the same order.
\par

The last stage of the envisaged experiment is the probing 
of the quantum state of the atomic ensembles that contain the clones,
in order to determine the experimentally attained fidelity. 
In the geometry shown in Fig.~1,
the pulse propagating along $z$ axis unavoidably interacts with both ensembles
$A$ and $B$. It is nevertheless possible to address the ensembles individually
by shining the light from a perpendicular direction, along the $y$ axis.
In this way one can measure the statistics of arbitrary quadrature and determine
the quadrature variances or even perform the full tomographic reconstruction of
the quantum state of the clones. Alternatively, atomic spins can be rotated 
between the measurements by applying pulses of magnetic field.
\par

In summary, we have proposed an experimentally feasible method of preparing 
long-lived atomic or atom-light clones for continuous quantum variables
of light. The protocols described here can be used in various quantum
communication protocols, e.g., for the optimal eavesdropping of a quantum key distribution scheme.
\par

We would like to thank F. Grosshans and S. Massar for stimulating discussions.
We acknowledge financial support from the Czech 
Ministry of Education under grant LN00A015,
from the Danish National Research Foundation (Danmarksgrundforskningsfond),
from the Communaut\'e Fran\c{c}aise de
Belgique under grant ARC 00/05-251, from the IUAP programme of the Belgian
government under grant V-18, and from the EU grants CAUAC, CHIC and SECOQC.

\end{document}